\documentstyle[12pt,doublespace,psfig]{article}

\def\ppap{
\mathrel{\hbox{\rlap{\hbox{\lower4pt\hbox{$\sim$}}}\hbox{$<$}}}}
\def\qqap{
\mathrel{\hbox{\rlap{\hbox{\lower4pt\hbox{$\sim$}}}\hbox{$>$}}}}

\newcounter{ref}%
\newcommand{\bib}{\refstepcounter{ref}{\scriptsize\arabic{ref}}}
\newcommand{\wbib}[1]{{\ref{#1}.}~}
\def\up#1{\hbox{${}^{\hbox{\sc #1}}$}}

\def\v@lp@ge:#1,#2:{{\bf#1},~#2}
\def\volumeandpage#1{\v@lp@ge:#1:}

\vspace*{-1cm}

\begin{document}

\centerline{\Large \bf A substantial amount of}
\centerline{\Large \bf hidden magnetic energy in the quiet Sun}
\vspace*{1cm}
\centerline{J.~Trujillo~Bueno$^{*,\dag}$, N.~Shchukina$^{+}$ \& A.~Asensio Ramos$^{*}$}
\vspace*{1cm}
\centerline{$^{*}$ Instituto de Astrof\'\i{}sica de Canarias,
             E-38205 La Laguna, Tenerife, Spain.}
\centerline{$^{+}$ Main Astronomical Observatory, National Academy of Sciences, 
Zabolotnogo 27, 03680 Kyiv, Ukraine}	    
\centerline{$^\dag$ Consejo Superior de Investigaciones Cient\'\i ficas, E-28006 Madrid, Spain.} 
\vspace*{1cm}
\centerline{{\sl e-mail address:} \ \ jtb@iac.es}

\vfil
\centerline{{\bf Nature} (2004; Vol. 430, 326--329)}

\vfil\eject
\noindent

{\bf 
Deciphering and understanding the small-scale magnetic activity 
of the quiet solar photosphere
should help to solve many of the key problems 
of solar and stellar physics, such as the magnetic
coupling to the outer atmosphere and the coronal heating\up{\bib\label{sten94},\bib\label{schrijver98},\bib\label{schrijver03}}.
At present, we can see only ${\sim}1\%$ 
of the complex magnetism of the quiet 
Sun\up{\scriptsize\ref{sten94},\bib\label{lin99},\bib\label{sa00},
\bib\label{cer03},\bib\label{khom03}}, which 
highlights the need to develop a reliable way to
investigate the remaining $99\%$. Here we report three-dimensional
radiative tranfer modelling of scattering polarization in
atomic and molecular lines that indicates the presence of
hidden, mixed-polarity fields on subresolution scales. Combining this
modelling with recent observational data\up{\bib\label{sten97},
\bib\label{jtb-them01},\bib\label{bom02},\bib\label{gandorfer}}
we find a ubiquitous tangled magnetic field
with an average strength of ${\sim}130$ G, which is much stronger in the
intergranular regions of solar surface convection than in the granular
regions. So the
average magnetic energy density in the quiet solar photosphere 
is at least two orders of magnitude greater than that derived
from simplistic one-dimensional investigations\up{\bib\label{fauro95},\bib\label{fauro01}},
and sufficient to balance radiative energy losses from the solar
chromosphere.
}

Most of our present empirical knowledge of solar surface magnetism
stems from the analysis of the light polarization that the Zeeman effect
induces in spectral lines\up{\scriptsize\ref{sten94},\scriptsize\ref{lin99},\scriptsize\ref{sa00},
\scriptsize\ref{cer03},\scriptsize\ref{khom03}}.
For example, the circular
polarization signals of which synoptic magnetograms are made
reveal the existence of an irregular network 
of spatially unresolved, intermittent flux patches
of kilogauss field concentrations
outlining the boundaries of the giant velocity cells of the 
supergranulation\up{\scriptsize\ref{sten94}}. 
Such routine magnetograms
give the (wrong) impression that the cell interiors 
(referred to here as internetwork regions) are non-magnetic,
and it is important to note that such regions
cover most of the solar `surface' at any given time
during the solar magnetic activity cycle.
However, high-spatial resolution magnetograms
show a multitude of mixed magnetic polarities
within the internetwork regions, with most of the detected
flux located in the intergranular lanes 
of the solar granulation pattern where the plasma
is downflowing\up{\scriptsize\ref{lin99},\scriptsize\ref{sa00},
\scriptsize\ref{cer03},\scriptsize\ref{khom03}}. 
The mean unsigned flux density 
turns out to be ${\sim}10$ G when choosing
the best compromise between
polarimetric sensitivity and spatio-temporal resolution
that is at present feasible. When such polarization signals
are interpreted taking into account the fact
that the magnetic field is not being
spatially resolved (for example, 
by assuming that one or more magnetic components
are coexisting with a `non-magnetic' component within the spatio-temporal 
resolution element of the observation), it is then found that the filling 
factor of the magnetic component(s) is ${\sim}1\%$.
The `problem' with the Zeeman effect is that the amplitudes of the measured
polarization signals are the smaller the greater the degree of
cancellation of mixed magnetic polarities within the
spatio-temporal resolution element of the observation.
Therefore, lack of detection does not necessarily imply
absence of magnetic fields. 

Fortunately, scattering  processes in spectral lines produce linear polarization signals, whose amplitudes are efficiently modified in the presence of tangled magnetic fields of strength $B_{\rm H}{\approx}1.137{\times}10^{-7}/(t_{\rm life}\,g_{\rm L})$ (with $B_H$ expressed in gauss, and where
$t_{\rm life}$ and $g_{\rm L}$ are respectively
the lifetime in seconds and Land\'e factor of the upper level of the transition). This so-called Hanle effect\up{\bib\label{hanle24},\bib\label{jtb01}} 
has the required diagnostic potential
for investigating spatially unresolved, `hidden' mixed-polarity 
magnetic fields in the solar atmosphere\up{\bib\label{sten82}}.
The problem is how to apply it to obtain reliable information given
that Hanle-effect diagnostics of such `turbulent' fields relies on a comparison
between the observed scattering polarization
and that corresponding to
the zero-field reference case. It has been pointed out correctly\up{\scriptsize\ref{fauro95},\scriptsize\ref{fauro01}}
that a suitable spectral line 
for Hanle-effect diagnostics of `turbulent' photospheric fields is that 
of Sr {\sc i} at 4607 \AA . However, the simplified approach
of assuming that the highly inhomogeneous and dynamic
solar photosphere can 
be well represented by a one-dimensional (1D) and static model is very
unreliable, because of the need to use the free parameters of `classical' stellar spectroscopy
(that is, micro- and macroturbulence for line broadening), which have a serious impact
on the calculated polarization amplitudes\up{\bib\label{shu-jtb03}}. 
We have shown\up{\scriptsize\ref{shu-jtb03}} 
that the particular 1D approach that has 
been applied\up{\scriptsize\ref{fauro95},\scriptsize\ref{fauro01}} 
yields artificially low values for the strength of the `turbulent' field
-that is, between 20 and 10 gauss as shown by
the dashed lines of Fig. 3 in Ref. {\scriptsize\ref{shu-jtb03}}. 

In order to improve the reliability of diagnostic
tools based on the Hanle effect, we have developed 
a novel approach based
on multilevel scattering polarization calculations in
three-dimensional (3D) models of the solar photosphere,
which we have obtained from realistic hydrodynamical
simulations of solar surface convection\up{\bib\label{asp00}}.
This has allowed us to obtain
the linear polarization amplitudes that scattering
processes in the inhomogeneous solar photosphere
would produce in the Sr {\sc i} 4607 \AA\ line
if there were no magnetic field. We point out that
our synthetic intensity profiles
(which take fully into account the Doppler shifts 
of the convective flow velocities in the 3D model)
are automatically in excellent agreement
with the observations when the meteoritic strontium 
abundance is chosen. However, 
we find that the calculated fractional linear 
polarization is substantially larger than the observed one, thus indicating
the need to invoke magnetic depolarization. 

As seen in Fig. 1, our radiative transfer simulations of 
the Hanle effect in the Sr {\sc i} line show
that a volume-filling, microturbulent magnetic field of about 60 gauss leads
to a notable agreement with the observed $Q/I$
(where $Q$ and $I$ are Stokes parameters).  In Fig. 1
there is a clear indication that the strength of the `turbulent' field 
required to explain the $Q/I$ observations
decreases with height in the atmosphere, from the 70 gauss
needed to explain the observations at $\mu=0.6$ to the 50 gauss
required to fit the observations
at $\mu=0.1$. This corresponds approximately to 
a height range between 200 and 400 km
above the solar visible `surface'. 

If the hidden magnetic field of the quiet solar photosphere
really had a strength of 60 G at
all points in the solar photosphere, then the mean magnetic
energy density ($E_m=<B^2>/8\pi$)
would be $E_m{\approx}140\,{\rm erg\,cm^{-3}}$,
which is an order of magnitude smaller than that
corresponding to the kG fields of the network patches. 
However, both magnetohydrodynamical simulations\up{\bib\label{cat99},\bib\label{sten-nord03}} and observations of the Zeeman effect\up{\scriptsize\ref{lin99},\bib\label{socas-sanchez},\scriptsize\ref{khom03}}
indicate that the photospheric
plasma of the quiet Sun has a continuous distribution of field
strengths, in such a way that the
weaker the field the larger the probability
of finding a magnetic strength between 
$B$ and $B+{\rm d}B$ when no distinction is made between
granular and intergranular 
regions. As shown by the black dashed-dotted  
line of Fig. 1, if we assume that the probability distribution function (PDF)
has an exponential shape (${\rm e}^{-B/B_0}/B_0$), we then find that $B_0\,{\approx}\,130$ 
gauss yields a fairly good fit to the observed fractional linear polarization. In this
much more realistic case $E_m{\approx}1300\,{\rm erg\,cm^{-3}}$
(that is, ${\sqrt{<B^2>}}\,{\approx}\,180$ G), which is about $20\%$ of the
averaged kinetic energy density produced by convective motions
at a height of 200 km in the 3D photospheric model. 
This result and the fact that the observed 
$Q/I$ does not seem to be modulated
by the solar cycle (see Fig. 1) suggests that a 
small-scale dynamo associated with turbulent motions
within a given convective domain of ionized 
gas\up{\scriptsize\ref{cat99},\scriptsize\ref{sten-nord03}} 
plays a significant role for the quiet Sun magnetism.
The total magnetic energy stored in the internetwork regions is now 
slightly larger than that corresponding to the kG fields of the network patches.
It is also of interest to point out that our empirical PDF 
with $<B>{\approx}\,\,130$ G produces a negligible Zeeman broadening 
of the emergent spectral line profiles, which is fully compatible with the reported 
empirical constraints\up{\scriptsize\ref{sten94}}. However, in contrast to what
might be expected from the conclusions of a recent paper\up{\bib\label{sanchez-emonet-cattaneo}}, the stretched exponential PDF that characterizes the surface distribution of magnetic fields `predicted' by the most recent numerical 
experiments of turbulent dynamos\up{\scriptsize\ref{cat99}},
which correspond to a magnetic Reynolds number of about 1000,
implies a Hanle-effect reduction of the scattering polarization amplitude of the 
Sr {\sc i} 4607 \AA\ line that is significantly 
smaller than what is needed to explain the observations. 

We do not have a sufficient number of observational constraints to
be able to determine the exact shape or the detailed spatial variation
of the solar PDF. However, we can conclude that 
the `turbulent' field is organized at the spatial scales of the solar granulation pattern, with relatively weak fields above the 
granules and with much stronger fields above the intergranular lanes
(see Fig. 2). As shown in Fig. 2 legend most of the volume
of the upflowing granules
where the observed C$_2$ line polarization originates
is occupied by magnetic fields much weaker than
what is needed to explain the observed depolarization
in the Sr {\sc i} 4607 \AA\ line. Therefore, most (but not all)
of the observed strontium line depolarization must be produced 
by relatively strong and 
tangled fields in the intergranular regions.

The contribution of the intergranular plasma to the observed
scattering polarization in the strontium line is very likely to
be close to the regime of Hanle saturation, which for this
spectral line occurs for magnetic strengths $B{\ge}300$ G.
This helps us to understand
why with low spatial and temporal resolution it is not easy 
(but definitely not impossible)
to detect spatial fluctuations in the observed $Q/I$ (Fig. 1). 
We have carried out calculations with several plausible PDFs,
assuming in all cases that the angular distribution
of the field vectors is isotropic and microturbulent 
for every single field strength. For instance, we have assumed an
exponential PDF (${\rm PDF}_{\rm G}=e^{-B/B_{\rm g}}/B_{\rm g}$) for the 
upflowing regions with a free parameter $B_{\rm g}$ to be determined
from the observations and a given
maxwellian function (${\rm PDF}_{\rm I}=2.38{\times}10^{-8}B^2{\rm exp}[-B/456]^2$) for the downflowing plasma obtained from
a best fit to the strong field part of the
intergranular histogram of the Zeeman splittings observed in near-IR
lines of neutral iron\up{\scriptsize\ref{khom03}}.
Note that this maxwellian PDF implies that the filling
factor of kG fields is ${\sim}2\%$ -that is, it does not exclude
the possibility\up{\scriptsize\ref{cer03}} of small-scale kG fields
in the internetwork regions. With this `educated' choice of plausible PDFs
the best agreement between theory and observations is
obtained for $B_{\rm g}{\approx}15$ G, which 
is consistent with the results of our
investigation of the Hanle effect in C$_2$ lines. 
We find that most of the magnetic energy turns out to be due to
rather chaotic fields in the intergranular plasma with
strengths between the equipartition field values and ${\sim}1$ kG. 
The total magnetic energy stored in the internetwork regions is now substantially larger than that corresponding to the kG fields of the supergranulation network.

Our conclusion that most of the volume of the  
granular regions is occupied by very weak fields seems to be compatible 
with the constraints imposed by our present understanding\up{\bib\label{jtb02}} 
of the `enigmatic' scattering polarization observed in the
Na {\sc i} D$_1$ line, whose line-core radiation
originates a few hundred kilometers 
higher in the solar atmosphere. As demonstrated
in Ref. {\scriptsize\ref{jtb02}},
the linear polarization signature of the scattered light
in the D$_1$ line is only completely suppressed in the presence of magnetic
fields larger than 10 G, regardless of their inclination. Note that this
result is significantly different to what had been concluded in an earlier
paper\up{\bib\label{landi98}}. In any case, it is
important to point out that the observed linear 
polarization in the sodium D$_1$ line still remains enigmatic because 
nobody has yet been able
to model both the amplitude and shape of the observed $Q/I$ 
profile\up{\bib\label{sten03}}.

Our empirical findings may have far-reaching implications 
in solar and stellar physics.
The hot outer regions of the solar atmosphere
(chromosphere and corona) radiate and expand, 
which takes energy. By far the largest energy
losses stem from chromospheric radiation with a
total energy flux of\up{\bib\label{anderson89}} 
${\sim}10^7$ ${\rm erg}\,\,{\rm cm}^{-2}\,\,{\rm s}^{-1}$.
Considering our most conservative estimate 
for the magnetic energy density -that is,
$E_m{\approx}140\,{\rm erg\,cm^{-3}}$- we obtain
an energy flux similar to the above-mentioned 
chromospheric energy losses when using either
the typical value of ${\sim}\,1\,{\rm km\,s^{-1}}$
for the convective velocities or
the Alfv\'en speed ($v_A=B/(4{\pi}{\rho})^{1/2}$, with $\rho$ the gas density). 
In reality, as pointed out above,
the true magnetic energy density that at any given time 
during the solar cycle is stored in 
the quiet solar photosphere is very much larger
than $140\,{\rm erg\,cm^{-3}}$. 
Only a relatively small fraction would thus suffice to
balance the energy losses of the solar outer atmosphere. 

Equally interesting is our result that $<B>{\approx}\,130$ G,
which indicates that the unsigned
magnetic flux density in the quiet solar photosphere is 
substantially large.
In fact, it has been suggested
recently that the dynamic geometry
of the magnetic connection between the photosphere
and the corona may sensitively depend on the amount of magnetic flux that
exists in the internetwork regions\up{\scriptsize\ref{schrijver03}}. 
The hidden flux that we have diagnosed turns out to be
organized at the spatial scales of the solar granulation pattern,
with much stronger fields in the intergranular regions.
Therefore, if the magnetic field of the 
`quiet' solar chromosphere is found to be
more complex than simply canopy-like, 
because of the influence of the
intergranular magnetic fields which are continuously moving around 
on temporal scales of minutes, then the
possibilities for magnetic reconnection\up{\bib\label{priest}} 
and energy dissipation
in the solar outer atmosphere would be much enhanced.
The `hidden' magnetic fields that we have 
reported here could thus provide the clue to understanding 
how the $10^6$ K solar corona is heated.

\baselineskip=.99\baselineskip \vfil\eject\noindent

\newpage

\noindent
\hrule

\noindent
{\bf References}

\footnotesize
\noindent
\begin{itemize}

\item[\wbib{sten94}] Stenflo, J.O.
   {\sl Solar Magnetic Fields: Polarized Radiation Diagnostics}.
   (Kluwer, Dordrecht, 1994)
   
\item[\wbib{schrijver98}] Schrijver, C.J. et al.
   {\sl Large-scale coronal heating by the small-scale
   magnetic field of the Sun}.
   {\sl Nature} {\bf 394}, 152 -- 154 (1998).   
   
\item[\wbib{schrijver03}] Schrijver, C.J. \& Title, A.
   The Magnetic Connection Between the Solar Photosphere and the Corona
   {\sl Astrophys. J.} {\bf 597}, L165-L168 (2003).   
      
\item[\wbib{lin99}] Lin, H. \& Rimmele, T.  
    The granular magnetic fields of the quiet Sun.
    {\sl Astrophys. J.} {\bf 514}, 448 -- 455 (1999).     
   
\item[\wbib{sa00}] S\'anchez Almeida, J. \& Lites, B.
   Physical properties of the solar magnetic photosphere under the MISMA
   hypothesis. II. Network and internetwork fields at the disk center.
   {\sl Astrophys. J.} {\bf 532}, 1215 -- 1229 (2000).    
    
\item[\wbib{cer03}] Dom\'\i nguez Cerde\~na, I., Kneer, F., \& S\'anchez Almeida, J.
	Quiet-Sun magnetic fields at high spatial resolution.    
	{\sl Astrophys. J.} {\bf 582}, L55 -- L58 (2003).  
		
\item[\wbib{khom03}] Khomenko, E., et al.
        Quiet-Sun internetwork magnetic fields observed in the infrared.
	{\sl Astron. Astrophys.} {\bf 408}, 1115-1135 (2003).  
	
\item[\wbib{sten97}] Stenflo, J.O., Bianda, M., Keller, C. \& Solanki, S.K.
   Center-to-limb variation of the second solar spectrum
   {\sl Astron. Astrophys.} {\bf 322}, 985 -- 994 (1997).
   
\item[\wbib{jtb-them01}] Trujillo Bueno, J., Collados, M., Paletou, F., \& Molodij, G.
   In {\sl Advanced Solar Polarimetry: Theory, Observations and Instrumentation} 
   (ed. Sigwarth, M.)
   141 -- 149
   (ASP Conf. Series Vol. 236, Astronomical Society of the Pacific, San Francisco, 2001). 
   
\item[\wbib{bom02}] Bommier, V., \& Molodij, G.
	Some THEMIS-MTR observations of the second solar spectrum (2000 campaign)
	{\sl Astron. Astrophys.} {\bf 381}, 241-252 (2002). 	
	
\item[\wbib{gandorfer}] Gandorfer, A.
The Second Solar Spectrum. Vol. 1: 4625 \AA\ to 6995 \AA\ , ISBN 3 7281 2764 7 (Z\"urich: vdf).	 
   
\item[\wbib{fauro95}] Faurobert-Scholl, M., et al.
	Turbulent magnetic fields in the solar photosphere:
	diagnostics and interpretation.
	{\sl Astron. Astrophys.} {\bf 298}, 289-302 (1995).  
	
\item[\wbib{fauro01}] Faurobert, M., Arnaud, J., 
        Vigneau, J. \& Frisch, H.
        Investigation of weak solar magnetic fields. 
	New observational results for the Sr {\sc i} 460.7 nm linear 
	polarization and radiative transfer modeling.
	{\sl Astron. Astrophys.} {\bf 378}, 627-634 (2001).	 
   
\item[\wbib{hanle24}] Hanle, W. 1924, \"Uber magnetische Beeinflussung der Polarisation
der Resonanzfluoreszenz, {\sl Z. Phys.} {\bf 30}, 93-105.   
   
\item[\wbib{jtb01}] Trujillo Bueno, J.
   In {\sl Advanced Solar Polarimetry: Theory, Observations and Instrumentation} 
   (ed. Sigwarth, M.).
   161--195
   (ASP Conf. Series Vol. 236, Astronomical Society of the Pacific, San Francisco, 2001).     
   
\item[\wbib{sten82}] Stenflo, J.O.
   The Hanle effect and the diagnostics of turbulent magnetic fields
   in the solar atmosphere.
   {\sl Solar Phys} {\bf 80}, 209 -- 226 (1982). 
   		
\item[\wbib{shu-jtb03}] Shchukina, N., \& Trujillo Bueno, J.
	In {\sl Solar Polarization 3} (eds. Trujillo Bueno, J. \& S\'anchez Almeida, J.)
	336 -- 343
        (ASP Conf. Series Vol. 307, Astronomical Society of the Pacific, San Francisco, 2003).         
        
\item[\wbib{asp00}] Asplund, M., Nordlund, \AA\, Trampedach, R., 
Allende Prieto, C. \& Stein, R. F.
	Line formation in solar granulation. I. Fe line shapes, shifts and
	asymmetries.
	{\sl Astron. Astrophys.} {\bf 359}, 729-742 (2000).  
	       
\item[\wbib{cat99}] Cattaneo, F.	 
	On the origin of magnetic fields in the quiet photosphere
	{\sl Astrophys. J.} {\bf 515}, L39--L42 (1999). 
              
\item[\wbib{sten-nord03}] Stein, R. F., \& Nordlund, \AA\ 
   In {\sl Modelling of Stellar Atmospheres} 
   (eds. Piskunov, N. E., Weiss, W. W. \& Gray, D. F.).
   169 -- 180 
   (ASP Conf. Series Vol. IAU 210, Astronomical Society of the Pacific, San Francisco, 2003).
   
\item[\wbib{socas-sanchez}] Socas-Navarro, H., \& S\'anchez Almeida, J.
   Magnetic fields in the quiet Sun: observational discrepancies and unresolved structure.
   {\sl Astrophys. J.} {\bf 593}, 581 -- 586 (2003).  
   
\item[\wbib{sanchez-emonet-cattaneo}] S\'anchez Almeida, J., Emonet, T., \& Cattaneo, F.
   Polarization of photospheric lines from turbulent dynamo simulations.
   {\sl Astrophys. J.} {\bf 585}, 536 -- 552 (2003).  
          
\item[\wbib{jtb02}] Trujillo Bueno, J., Casini, R., Landolfi, M. \& Landi Degl'Innocenti, E.  
    The physical origin of the scattering polarization of the Na {\sc i} D lines
    in the presence of weak magnetic fields
    {\sl Astrophys. J.} {\bf 566}, L53 -- L57 (2002).  
    
\item[\wbib{landi98}] Landi Degl'Innocenti, E. 
    Evidence against turbulent and canopy-like magnetic fields 
    in the solar chromosphere
    {\sl Nature} {\bf 392}, 256 -- 258 (1998).     
    
\item[\wbib{sten03}] Stenflo, J. O.
	In {\sl Solar Polarization 3} (eds. Trujillo Bueno, J. \& 
	S\'anchez Almeida, J.)
	385 -- 398
        (ASP Conf. Series Vol. 307, Astronomical Society of the Pacific, San Francisco, 2003).       
    
\item[\wbib{anderson89}] Anderson, L.S. \& Athay, R.G.	 
	Model solar chromosphere with prescribed heating
	{\sl Astrophys. J.} {\bf 346}, 1010--1018 (1989). 
  
\item[\wbib{priest}] Priest, E. \& Forbes, T.
   {\sl Magnetic Reconnection: MHD Theory and Applications}.
   (Cambridge University Press, New York, 2000)       
   	   
\item[\wbib{landi83}] Landi Degl'Innocenti, E. 
    Polarization in Spectral Lines: I. A Unifying Theoretical Approach.
    {\sl Solar Phys.} {\bf 85}, 3 -- 31 (1983).  
    
\item[\wbib{jtb03a}] Trujillo Bueno, J.
   In {\sl Stellar Atmosphere Modeling} 
   (eds. Hubeny, I., Mihalas, D. \& Werner, K.).
   551--582
   (ASP Conf. Series Vol. 288, Astronomical Society of the Pacific, San Francisco, 2003).        
    
\item[\wbib{jtb03b}] Trujillo Bueno, J.
	In {\sl Solar Polarization 3} (eds. Trujillo Bueno, J. \& 
	S\'anchez Almeida, J.)
	407 -- 424
        (ASP Conf. Series Vol. 307, Astronomical Society of the Pacific, San Francisco, 2003).                             
   
\end{itemize}   

%{\bf Acknowledgements} 
We thank F. Kneer, E. Landi Degl'Innocenti and
F. Moreno-Insertis
for scientific discussions.
We are also grateful to P. Fabiani Bendicho 
for help with the numerical solution 
of the 3D radiative transfer equation.
This research has been supported by the
Spanish Plan Nacional de Astronom\'\i a y Astrof\'\i sica
and by the European Commission via the INTAS
program and the Solar Magnetism Network.

Correspondence and requests for materials should be addressed
to J.T.B. (e-mail: jtb@iac.es).

\vfill
\eject

\normalsize

{\bf Figure 1. Spectropolarimetric observations 
versus 3D modelling of the Hanle effect.}

This figure shows the center-to-limb variation 
of the fractional linear polarization
at the core of the Sr {\sc i} $4607$ \AA\ line after substraction of the
continuum polarization level. (The Stokes $I$ and $Q$ parameters are
defined in Ref. [1]; $\mu={\rm cos}\,{\theta}$, with $\theta$ the angle
between the solar radius vector through the observed point and the line of sight). Open circles, various observations\up{\scriptsize\ref{sten97}} taken 
during a minimum period of the solar cycle (in particular,
during September--October 1995). 
The remaining symbols correspond to observations
taken during the most recent maximum period 
of the solar cycle. Diamonds, 
observations\up{\scriptsize\ref{jtb-them01}} obtained
during May 2000; `crosses' and `plus'
symbols, observations
taken during August 2000 (see Ref. {\scriptsize\ref{bom02}}) 
and December 2002 (V. Bommier 2002; personal communication), 
respectively; filled circles,
observations obtained during September 2003
at the Istituto Ricerche Solari Locarno (Switzerland)
in collaboration with M. Bianda. Coloured lines, the results of
our 3D scattering polarization calculations in the presence  
of a volume-filling and single-valued microturbulent
field with an isotropic distribution of directions
below the mean free path of the line photons
(from top to bottom: 0, 5, 10, 15, 20, 30, 
40, 50, 60, 80, 100, 150, 200, 250 and 300 gauss).
We have solved the relevant equations\up{\bib\label{landi83}} 
via the application of efficient radiative transfer 
methods\up{\bib\label{jtb03a}}, and
using realistic collisional depolarizing
rate values\up{\scriptsize\ref{fauro95}}, 
which turn out to be the largest rates among
those found in the literature. Note that there is no evidence of a serious modulation
of the strength of the `turbulent' field with the last solar activity cycle, and 
that the best average fit to the observations is obtained for 60 G.
The black dashed-dotted line indicates the resulting $Q/I$ line-core amplitudes
for the case of a single exponential PDF (${\rm e}^{-B/B_0}/B_0$) with $B_0=130$ G, which implies $<B>\,=\,\int{B\,{\rm PDF}(B)\,dB\,}=130$ G.

\vfill
\eject

{\bf Figure 2. Evidence for a `turbulent' field organized at the
spatial scales of the solar granulation pattern.} 

This figure demonstrates that the
scattering polarization observed in molecular lines\up{\scriptsize\ref{gandorfer}}
comes mainly from the upflowing regions of the `quiet' solar photosphere.
The solid line contours delineate such upflowing
regions at two heights in the 3D photospheric model,
which indicate the approximate atmospheric region
where the observed C$_2$ line radiation originates. 
The two top panels show the horizontal fluctuation of the 
`degree of anisotropy' (${\cal A}={J^2_0/J^0_0}$)\up{\scriptsize\ref{landi83},\scriptsize\ref{jtb01}}
of the solar continuum radiation at 5000 \AA\ .
(Note that there is a very strong correlation between the
upflowing regions and ${\cal A}$). The two bottom panels visualize the corresponding horizontal fluctuation in the number density (${\cal N}$) of C$_2$ molecules.
(Note that there is a significant correlation between the
upflowing regions and ${\cal N}$). Since in the solar atmosphere
the scattering polarization in weak molecular lines
is proportional to both ${\cal A}$ and ${\cal N}/{\eta_c}$
(where ${\eta_c}$ is the background continuum opacity, which
is anticorrelated with ${\cal N}$ in such range of heights)
we conclude that only the upflowing regions make a significant
contribution to the observed molecular 
scattering polarization. Therefore, we can obtain 
information on the distribution of magnetic fields in such (granular)
upflowing regions of the solar photosphere via theoretical modelling of the
observed scattering polarization in suitably chosen pairs of C$_2$ lines\up{\bib\label{jtb03b}}. 
We have calculated the magnetic sensitivity of the observed C$_2$ lines by
taking into account all the relevant optical pumping mechanisms in a very realistic multilevel model for C$_2$. Thanks to this Hanle-effect investigation,
we have been able to conclude that in the granular regions
the `turbulent' field is much weaker
than what is needed to explain the observed depolarization in the
Sr {\sc i} 4607 \AA\ line. For this reason, the intergranular regions must
be pervaded by relatively strong tangled fields capable 
of producing most of the observed depolarization in the strontium line.

\vfill
\eject

\begin{figure}
\psfig{figure=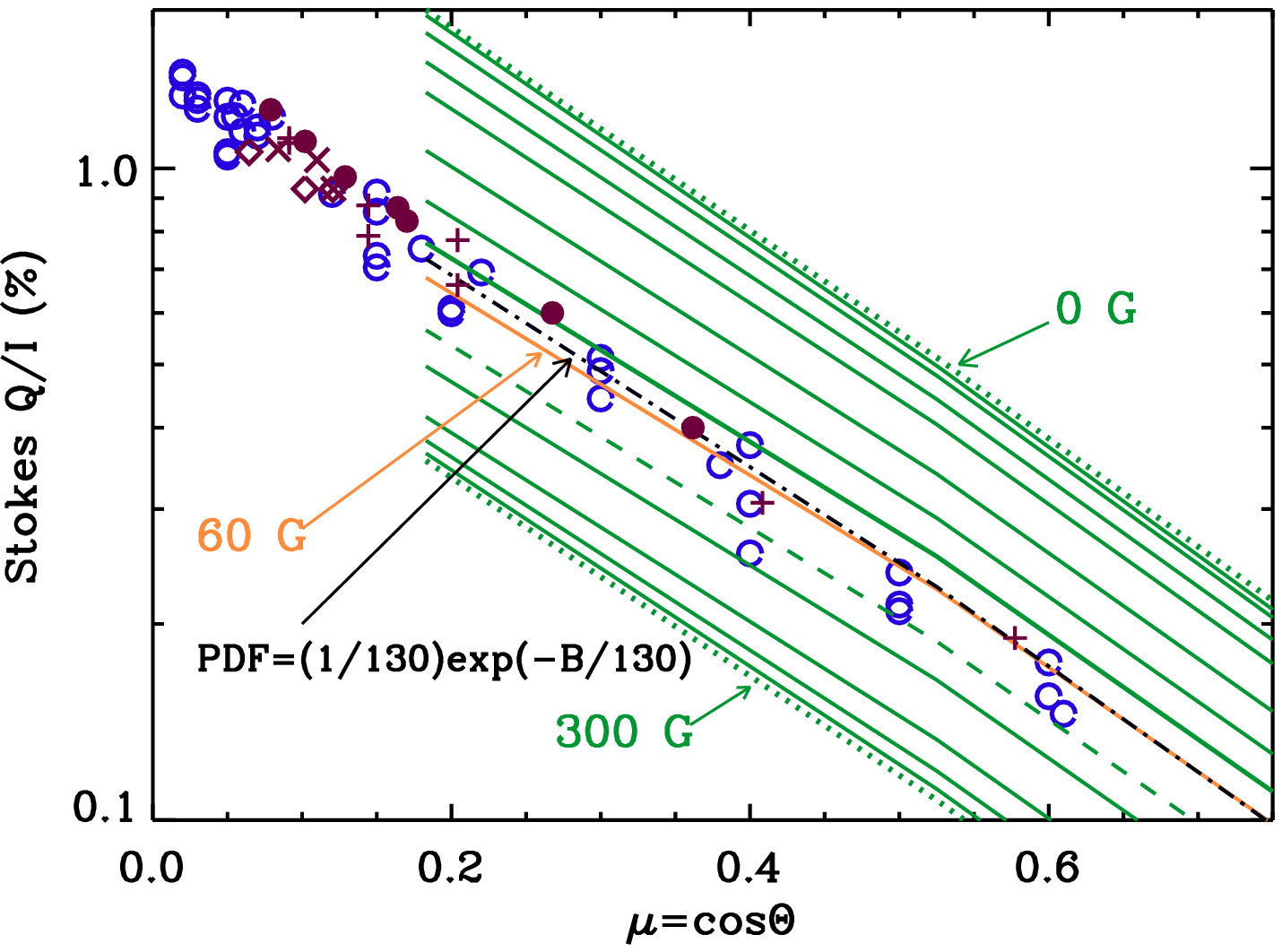}
\end{figure}

{\bf Figure 1}

\vfill
\eject

\begin{figure}
\psfig{figure=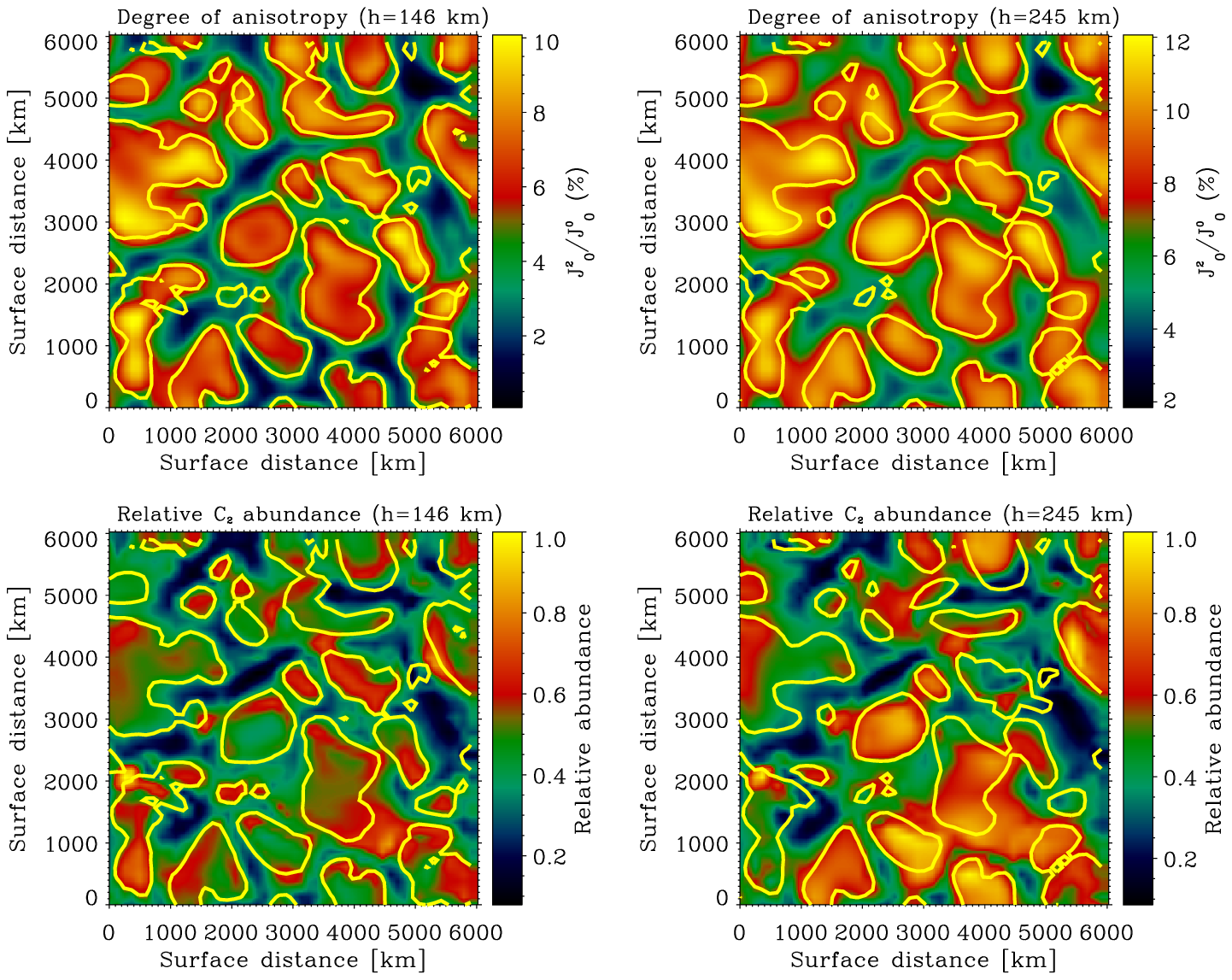}
\end{figure}

{\bf Figure 2}

\end{document}